\documentclass[12pt]{report}	% The documentclass must be ``report''.

\usepackage{utdiss2}  	
\usepackage{amsmath,amsthm,amsfonts,amscd}  % Some packages to write mathematics.
\usepackage{eucal} 	 	% Euler fonts
\usepackage{verbatim}      	% Allows quoting source with commands.
\usepackage{makeidx}       	% Package to make an index.
\usepackage{epsfig}         	% Allows inclusion of eps files.
\usepackage{listings}
\usepackage{morefloats}
\usepackage{color}
\usepackage{courier}
\usepackage{cite}
\usepackage{graphicx,amssymb,amstext,amsmath,amsthm,graphics,multicol}
\author{Ahmed Abdelhadi} 
\coauthor{Haya Shajaiah}  	% Required
\title{Optimal Resource Allocation for Cellular Networks with MATLAB Instructions}% Required
\theoremstyle{definition}

\theoremstyle{remark}
\newtheorem{rem}{Remark}[section]

\newcommand{\latexe}{{\LaTeX\kern.125em2%
                      \lower.5ex\hbox{$\varepsilon$}}}

\chardef\bslash=`\\	% \bslash makes a backslash (in tt fonts)
\makeatletter		% Starts section where @ is considered a letter
\def\square{\RIfM@\bgroup\else$\bgroup\aftergroup$\fi
  \vcenter{\hrule\hbox{\vrule\@height.6em\kern.6em\vrule}%
                                              \hrule}\egroup}
\makeatother		% Ends sections where @ is considered a letter.
\makeindex    % Make the index

\begin{document}

\titlepage              % Produces the title page.
%%%%%%%%%%%%%%%%%%%%%%%%%%%%%%%%%%%%%%%%%%%%%%%%%%%%%%%%%%%%%%%%%%%%%%
\tableofcontents   % Table of Contents will be automatically
                   % generated and placed here.
\listoftables      % List of Tables and List of Figures will be placed
\listoffigures     % here, if applicable.

%%%%%%%%%%%%%%%%%%%%%%%%%%%%%%%%%%%%%%%%%%%%%%%%%%%%%%%%%%%%%%%%%%%%%%
% Actual text starts here.					     %
%%%%%%%%%%%%%%%%%%%%%%%%%%%%%%%%%%%%%%%%%%%%%%%%%%%%%%%%%%%%%%%%%%%%%%
\singlespacing
\setlength\parindent{0pt}

\definecolor{dkgreen}{rgb}{0,0.6,0}
\definecolor{gray}{rgb}{0.5,0.5,0.5}

\lstset{language=Matlab,
   keywords={break,case,catch,continue,else,elseif,end,for,function,
      global,if,otherwise,persistent,return,switch,try,while},
   basicstyle=\ttfamily,
   keywordstyle=\color{blue},
   commentstyle=\color{dkgreen},
   stringstyle=\color{red},
   numbers=left,
   numberstyle=\tiny\color{gray},
   stepnumber=1,
   numbersep=10pt,
   backgroundcolor=\color{white},
   tabsize=4,
   showspaces=false,
   breaklines=true,
   showstringspaces=false}
   
%\doublespacing
%
\chapter{Introduction} \label{Ch:1}

This report presents a more detailed description of the algorithm and simulations published in papers \cite{AbdelhadiICNC2014, AbdelhadiPIMRC2013}. It includes a step by step description of the algorithm and included the corresponding flow chart. In addition, detailed instructions of the MATLAB code used to simulate the proposed allocation algorithm in \cite{AbdelhadiICNC2014, AbdelhadiPIMRC2013} is presented. The report starts with a brief motivation of resource allocation problem in wireless networks. Then, some of the prior related work on the subject are mentioned. Finally, we provide the details instructions on MATLAB functions used in our algorithm. More rigorous analysis and proofs of the problem and algorithm are present in \cite{AbdelhadiICNC2014, AbdelhadiPIMRC2013} and further discussions presented in \cite{Ghorbonzadeh_book1_chapter3, Ghorbonzadeh_book1_chapter4}.

\section{Motivation and Background}

There is a significant increase in the number of users and volume of traffic for wireless services \cite{Infonetics, Nokia2013, Zokem2011, Nokia2011}. Hence, it urges for improvement of quality of experience (QoE) \cite{QoE_paper}, sometimes called quality of service (QoS) in some articles \cite{QoS_3GPP, Ekstrom2006, Ghorbonzadeh_book1_chapter1}, of cellular networks \cite{Ghosh2011}. This improvement needs to be conducted on multiple layers of the network. Some progress to enhance the service in link layer has been conducted in \cite{Piro2010, Monghal2008, Soldani2011, Hujun2006}. Other researchers advanced the user experience by conducting design improvement to physical layer as shown in \cite{Larmo2009, ciochina2010}. The utilization of game theory methods in \cite{Ali2011, Fudenberg1991} and microeconomics in \cite{Ranjan2011, Johari2011} provided improvement to QoE.

Relaying on network layer QoS research was conducted by \cite{Chung2010} with consideration to energy efficiency. QoS was studied within the context of LTE third generation partnership project (3GPP) standardization \cite{3GPP, phy_channels_modulation, phy_procedures} in \cite{Le2013conf, Le2013} and within WiMAX \cite{Andrews2007} in \cite{Alasti2010, Niyato2007}. QoS at the Network layer was exploited from policy management perspective in \cite{IXIACOM2010} for Mobile Broadband \cite{FCC2010} and in \cite{Li2003} for Universal Mobile Terrestrial System (UMTS) \cite{ETSI2012, ETSI2016}. End to end QoS was proposed in \cite{Ahmed2004} and component-based QoS was proposed in \cite{Tournier2005}. Another method for improving QoS is via hardware by increasing the battery life which was discussed in \cite{Jung2011, Tellabs2012}. A solution that supports real-time traffic is proposed in \cite{Gorbil2011}.

For operators to deliver better service to their customers, QoS needs to be address efficiently via cross-layer design. Some researchers suggested global coordination between layers of Open Systems Interconnection (OSI) model \cite{Stallings2013} as in \cite{Dovrolis2002, Sali2005}. Other researchers modified the Asynchronous Transfer Mode (ATM) network protocol stack to achieve cross-layer QoS as in \cite{Lutz1991, Perros1994}. One the other hand, application layer QoS was the focus of the study in \cite{Kbah_ICNC2016, Erpek_SysCon_2016}. 

For wired IP networks, Integrated Services (IntServ) and Differentiated Services (DiffServ) were proposed in \cite{Braden1994, Braden1997} and  \cite{Blake1998, Nichols1999, Nahrstedt1995}, respectively. These methods focus on QoS on the routers in the form of scheduling, routing and shaping. 

In dealing with resource allocation various formulations are adapted, e.g. proportional fairness \cite{Kushner2004, Andrews2001, UtilityFairness} and max-min fairness \cite{Li2011, Prabhu2010, DBLP:conf/qosip/Harks05, Nandagopal2000}, as they achieve optimality for inelastic traffic \cite{Fudenberg1991, Ghorbonzadeh_book1_chapter2}. Network proportional fairness models were proposed with optimal solution for elastic traffic in \cite{kelly98ratecontrol, Low99optimizationflow} and weighted fair queuing (WFQ) in \cite{Parekh1993,Demers1989}. Some attempts to extend to inelastic traffic was conducted in \cite{RebeccaThesis}. However, optimality was shown in \cite{abdelhadi2015optimal, Abdelhadi_context} using convex optimization techniques \cite{Boyd} and the sensitivity to traffic is shown in \cite{Ghorbonzadeh_book1_chapter5}. Multi-class service offering with real-time application, using sigmoid functions was shown in \cite{Lee05non-convexoptimization, DL_PowerAllocation, Abdelhadi_MobiCom2013}. Extension to include resource blocks were developed in \cite{GhorbanzadehICNC2015, Erpek2015, Ghorbonzadeh_book1_chapter6}.

The President's Council of Advisers on Science and Technology (PCAST)’s report \cite{PCAST2012} recommends the use of the government-held spectrum to expand the available spectrum for mobile communications and so increase the service quality and meet future demands as well \cite{Wilson2010}. As a result, Federal Communications Commission (FCC) is studying sharing of under-utilized spectrum, e.g. S-band radars \cite{Richards2010, FCC_5GHz_Radar06}, with over-utilized spectrum \cite{FCC2012, FCC_NBP10} and the National Telecommunications and Information Administration (NTIA) is studying the effect of interference between mobile broadband systems and other wireless systems, e.g. WiMAX and radar \cite{NTIA2010, NTIA12, NTIA2014}. 

A non-convex optimization approaches to maximize system utilities for the case of multiple carriers were proposed in \cite{DBLP:conf/globecom/TychogiorgosGL11, DBLP:conf/pimrc/TychogiorgosGL11, Tao2008, Yuan10CarrierAggregation, ResourceAllocationConsideration} followed by convex optimization approaches in \cite{AbdelhadiICNC2015, Shajaiah_Springer2015}. The aggregation of radar spectrum to cellular spectrum was presented in \cite{GhorbanzadehMILCOM2014, Abdelhadi_ICNC_2016_network_MIMO, Ghorbonzadeh_book1_chapter7} to provide solutions for the spectrum sharing problem presented in \cite{Lackpour2011, Sanders2013, FHH14, Hossain_radar1}. 

The resource allocation solution proposed in \cite{AbdelhadiICNC2014, AbdelhadiPIMRC2013} is generic and can be applied to many systems, e.g. multi-cast network \cite{Abdelhadi_ITW2010}, ad-hoc network \cite{AnewDistributedOptimization, Jose_INFOCOM2010} and WiFi network \cite{Abdelhadi_IEEE_SysCon2016, Hossain_WiFi1, Hossain_WiFi2}. Some successful usages of that solution for machine to machine (M2M) communications were conducted in \cite{Kumar_M2M1, Kumar_M2M2, Kumar_M2M3, Kumar_M2M4, Kumar_M2M6} where optimization is with latency constraints rather than bandwidth constraints.

\chapter{User Applications Utilities}
%\pagenumbering{arabic}
\noindent The user satisfaction with the provided service can be expressed using utility functions that represent the degree of satisfaction of the user function of the rate allocated by the cellular network \cite{AbdelhadiICNC2014, AbdelhadiPIMRC2013}. We assume that the applications utility functions $U(r)$ are strictly concave or sigmoid functions. 
%%%%%%%%%%%%%%%%%%%%%%%%
\section{Sigmoid Utility}
The normalized sigmoid utility function is used in this cellular system, as in \cite{DL_PowerAllocation, Wang_SysCon2016}. It can be expressed as 
\begin{equation}\label{eqn:sigmoid}
U(r) = c\Big(\frac{1}{1+e^{-a(r-b)}}-d\Big)
\end{equation}
where $c = \frac{1+e^{ab}}{e^{ab}}$ and $d = \frac{1}{1+e^{ab}}$. So, it satisfies $U(0)=0$ and $U(\infty)=1$. The inflection point of normalized sigmoid function is at $r^{\text{inf}}=b$.

In MATLAB, the sigmoid utility code is
 \begin{lstlisting}
y(i) = c(i).*(1./(1+exp(-a(i).*(x-b(i))))-d(i));

\end{lstlisting}
where 
 \begin{lstlisting}
c = (1+exp(a.*b))./(exp(a.*b));
d = 1./(1+exp(a.*b));

\end{lstlisting}
%%%%%%%%%%%%%%%%%%%%%%%%%%
\section{Logarithmic Utility}
The normalized logarithmic utility function is used as well, as in \cite{Wilson1993, UtilityFairness, Shenker95fundamentaldesign}, that can be expressed as 
\begin{equation}\label{eqn:log}
U(r) = \frac{\log(1+kr)}{\log(1+kr^{\text{max}})}
\end{equation}
where $r^{\text{max}}$ is the rate achieving 100\% user satisfaction and $k$ is the rate of increase with rate $r$. So, it satisfies $U(0)=0$ and $U(r^{\text{max}})=1$. The inflection point of normalized logarithmic function is at $r^{\text{inf}}=0$. 

In MATLAB, the logarithmic utility code is
 \begin{lstlisting}
y2(i) = log(k(i).*x+1)./(log(k(i).*100+1));

\end{lstlisting}

\begin{figure}[t]
\centering
  \includegraphics[scale=0.4]{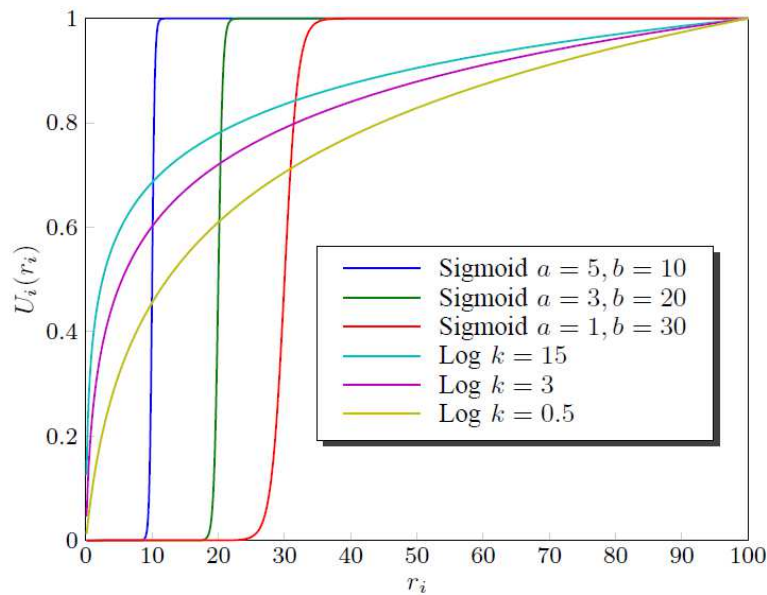}
  \caption{Applications Utilities}
  \label{fig:app_utilities}
\end{figure}

The utility functions with the parameters in Table \ref{tab:utility} are shown in Figure \ref{fig:app_utilities} \cite{AbdelhadiICNC2014, Wang_ICNC2016}.
%%%%%%%%%%%%%%%%%%%%%%%%%%%%%%%%%%
\begin {table}[h]
\caption {Applications Utilities} 
\label{tab:utility} 
\begin{center}
\begin{tabular}{| p{1cm} | p{3cm} | p{4.5cm}|| p{1cm} | p{3.5cm} |}
%\label{table:utility}
  \hline
  Sig1 & $a=5,\:\: b=10$ & e.g. VoIP & Log1 & $k=15,\:\:r_{max}=100$  \\ \hline
  Sig2 & $a=3,\:\: b=20$ & e.g. SD video streaming & Log2 & $k=3,\:\:r_{max}=100$ \\ \hline
  Sig3 & $a=1,\:\: b=30$ & e.g. HD video streaming & Log3 & $k=0.5,\:\:r_{max}=100$ \\ \hline
\end{tabular}
%\caption {Should be a caption}
\end{center}
\end {table}
%%%%%%%%%%%%%%%%%%%%%%%%%%%%%%%%%%
%%%%%%%%%%%%%%%%%%%%%%%%%%
%%%%%%%%%%%%%%%%%%%%%%%%%%
\section{Utilities used in Simulation}
We use three normalized sigmoid function that are expressed by equation (\ref{eqn:sigmoid}) with different parameters:
\begin{itemize}
 \item $a = 5$, $b=10$ which is an approximation to a step function at rate $r =10$ (e.g. VoIP), 
 \item $a = 3$, $b=20$ which is an approximation of an adaptive real-time application with inflection point at rate $r=20$ (e.g. standard definition video streaming)
  \item $a = 1$,  $b=30$ which is also an approximation of an adaptive real-time application with inflection point at rate $r=30$ (e.g. high definition video streaming). 
\end{itemize}
We use three logarithmic functions that are expressed by equation (\ref{eqn:log}) with $r_{max}$ =100 and different $k_i$ parameters which are approximations for delay tolerant applications (e.g. FTP). We use $k =\{15, 3, 0.5\}$. 

In MATLAB, the code for plotting the utilities and their derivatives code is
 \begin{lstlisting} 
function utility_fn
close all
clear all
clc
syms x
%x = 0:0.1:100;
k = [15 3 0.5];
a = [5 3 1];
b = [10 20 30];
c = (1+exp(a.*b))./(exp(a.*b));
d = 1./(1+exp(a.*b));
%x = zeros(1,1000)
for i = 1: length(a)
    y(i) = c(i).*(1./(1+exp(-a(i).*(x-b(i))))-d(i));
    %y2(i) = k(i).*log((exp(1).*x+1)./x);
    %y2(i) = a(i).*log(b(i).*x+1)./(1+ a(i).*log(b(i).*100+1));
    y2(i) = log(k(i).*x+1)./(log(k(i).*100+1));
end
z = log(y);
z2 = log(y2);
for i = 3: 4%length(a)
    for j = 1:1: 101 %j = 1: 1000
        x0(j) = 1 * j -1 %% x0(j) = 0.1 * j;
        yy(j,i) = subs(y(i),x0(j));
        yy2(j,i) = subs(y2(i),x0(j));
        dy(j,i) = diff(y(i),x);
        dy2(j,i) = diff(y2(i),x);
        dyy(j,i) = subs(dy(j,i),x,x0(j));
        dyy2(j,i) = subs(dy2(j,i),x,x0(j));
        ddy(j,i) = diff(dy(i),x);
        ddy2(j,i) = diff(dy2(i),x);
        ddyy(j,i) = subs(ddy(j,i),x,x0(j));
        ddyy2(j,i) = subs(ddy2(j,i),x,x0(j));
        zz(j,i) = subs(z(i),x0(j));
        zz2(j,i) = subs(z2(i),x0(j));
        dz(j,i) = diff(z(i),x);
        dz2(j,i) = diff(z2(i),x);
        dzz(j,i) = subs(dz(j,i),x,x0(j));
        dzz2(j,i) = subs(dz2(j,i),x,x0(j));
    end
end
%dy = double(dyy)
subplot(3,1,1); 
plot(x0,yy,x0,yy2)

subplot(3,1,2); 
plot(x0,dyy,x0,dyy2)

subplot(3,1,3); 
plot(x0,ddyy,x0,ddyy2)

figure;
subplot(2,1,1); 
plot(x0,zz,x0,zz2)

subplot(2,1,2); 
plot(x0,dzz,x0,dzz2)
%z = log(y);
%z2 = log(y2);
%plot(x,z,x,z2)

\end{lstlisting}
\section{Implementation Example}
In this example we use utility functions for youtube and FTP file transfer. Empirically, it was found that below 200 kbps, youtube crashed and buffered constantly \cite{Ghorbanzadeh_ICNC_2015}.  Above 740 kbps there was negligible gain. So for your example, a rough estimate would be to use a sigmoid-like utility where
\begin{itemize}
 \item 200 kbps == 5\% satisfaction (or could be something between 1-10\%).
\item 740 kbps == 99\% satisfaction.
\end{itemize}
with inflection point $\frac{(740+200)}{2} = 470$ kbps (i.e. $b$ = 470 kbps) and the slope is $\frac{(99-5)}{(740-200)} = 0.174$ \%per kbps (i.e. $a$ = 0.174).

\chapter{Single Carrier with Single Utility per User}\label{sec:single_utility}
%%%%%%%%%%%%%%%%%%%%%%%%%
 \section{Optimal Resource Allocation}\label{sec:optimal}
\noindent 

%The material in this code are from the following references \cite{ShajaiahICNC2014,ShajaiahMILCOM2013, ShajaiahPIMRC2014,Shajaiah_ARXIV_1503.08994}
%%%%%%%%%%%%%%%%%%%%%%%%%
\subsection{System Model of Single Carrier with Single Utility per User}\label{subsec:modelsingle}
\begin{figure}[h]
  \centering
    \includegraphics[scale=0.8]{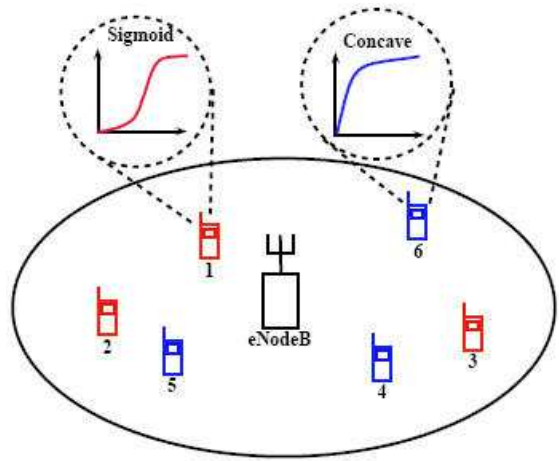}
      \caption{System Model of Single Carrier with Single Utility per User}
      \label{fig:model1}
\end{figure}
In our simulation, we consider a single cell in a mobile network consisting of a single base station and $M$ users ($M=6$ shown in Figure \ref{fig:model1}). The bandwidth allocated by the base station to $i^{th}$ user is given by $r_i$. Each user has its own utility function $U_i(r_i)$ that corresponds to the type of traffic being handled by the user. Our objective in this report, stated more rigorously in \cite{AbdelhadiICNC2014,AbdelhadiPIMRC2013}, is to determine the bandwidth the base station should allocate to the users. We assume the utility functions $U_i(r_i)$ to be strictly concave or sigmoid functions. 
%%%%%%%%%%%%%%%%%%%%%%%%%
\subsection{Algorithm of Optimal Resource Allocation}
  \begin{figure}[h]
  \centering
    \includegraphics[scale=1]{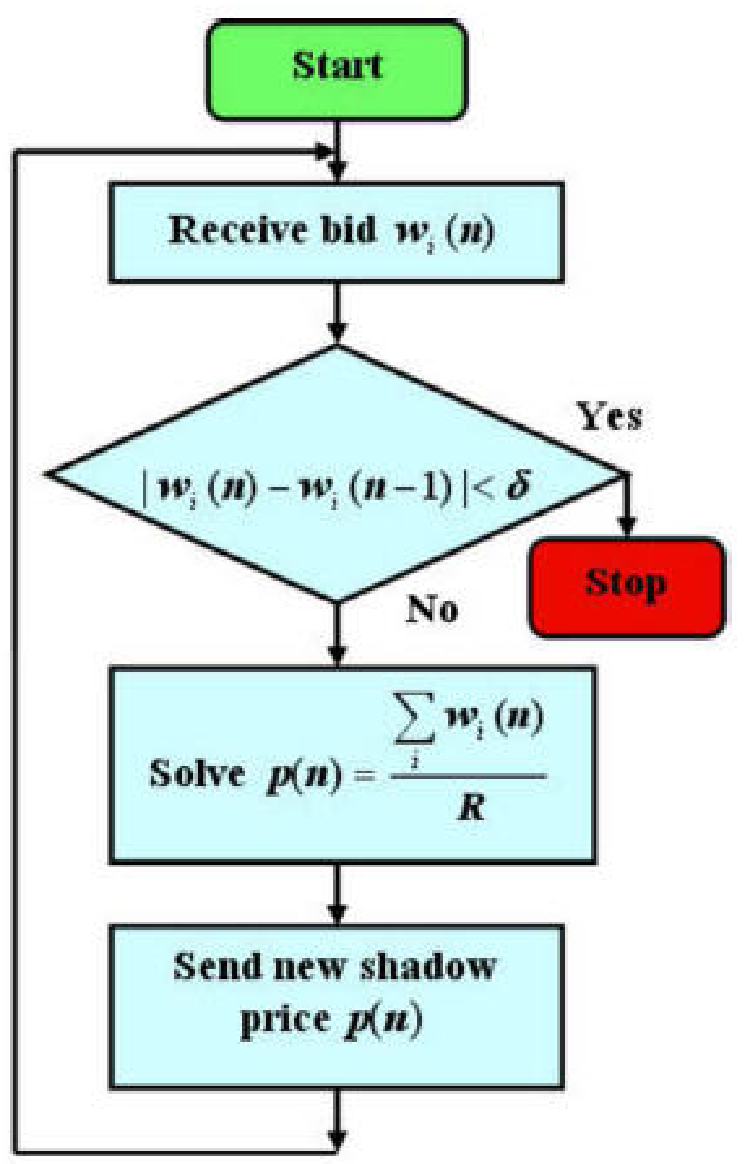}
      \caption{Base Station Algorithm of Single Carrier with Single Utility per User}
      \label{fig:base stationalgoptimal}
\end{figure}
%%%%%%%%%%%%%%%%%%%%%%%
The distributed resource allocation algorithm for of single carrier cell with users with single utility. It is an iterative solution for allocating the network resources with utility proportional fairness.
%%%%%%%%%%%%%%%%%%%%%
  \begin{figure}[h]
  \centering
    \includegraphics[scale=1]{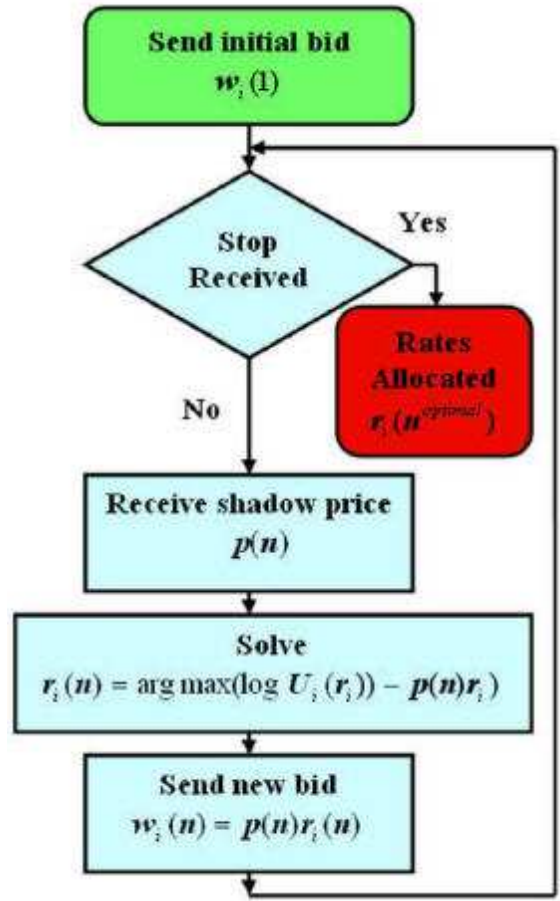}
      \caption{User Algorithm of Single Carrier with Single Utility per User}
      \label{fig:UEalgoptimal}
\end{figure}
The algorithm is divided into an user algorithm shown in flow chart in Figure \ref{fig:UEalgoptimal} and an base station algorithm shown in flow chart in Figure \ref{fig:base stationalgoptimal}.  Flow Chart Description:
\begin{itemize}
 \item Each user starts with an initial bid $w_i(1)$ which is transmitted to the base station. 
 
 In MATLAB
 \begin{lstlisting} 
% Initial Bids
w = [10 10 10 10 10 10];

\end{lstlisting}
 \item The base station calculates the difference between the received bid $w_i(n)$ and the previously received bid $w_i(n-1)$ and exits if it is less than a pre-specified threshold $\delta$. 
 
 In MATLAB
\begin{lstlisting}
while  (delta > 0.001) %(time<80)%(
    :
    :
    :
    :
    delta = max(abs(w-w_old))
end

\end{lstlisting}
 \item We set $w_i(0) = 0$. If the value is greater than the threshold $\delta$, base station calculates the shadow price $p(n) = \frac{\sum_{i=1}^{M}w_i(n)}{R}$ and sends that value to all the users. 
 
 In MATLAB
\begin{lstlisting}
function [p] = base station(w,Rate)
R = Rate;
p = sum(w)/R;

\end{lstlisting}
 \item Each user receives the shadow price to solve for the rate $r_i$ that maximizes $\log U_i(r_i) - p(n)r_i$.  
 
 In MATLAB
 \begin{lstlisting} 
for i = 1: length(a)
    y(i) = log(c(i).*(1./(1+exp(-a(i).*(x-b(i))))-d(i)));
    y(length(a)+i) = log(log(k(i).*x+1)./(1+ log(k(i).*100+1)));
end
for i = 1: 2*length(a)
    dy(i) = diff(y(i),x);
end
:
:
S(i) = dy(i)-p(time);
soln(i,:) = double(solve(S(i)));
:
:

\end{lstlisting}
 \item That rate is used to calculate the new bid $w_i(n)=p(n) r_{i}(n)$. 
 
 In MATLAB
 \begin{lstlisting}
w(i) = r_opt(i) * p(time);

\end{lstlisting}
 \item Each user sends the value of its new bid $w_i(n)$ to the base station. This process is repeated until $|w_i(n) - w_i(n-1)|$ is less than the pre-specified threshold $\delta$. 
 
 In MATLAB
 \begin{lstlisting}
while  (delta > 0.001) %(time<80)%(
    :
    :
    :
    :
    delta = max(abs(w-w_old))
end

\end{lstlisting}
\end{itemize}
%%%%%%%%%%%%%%%%%%%%
\begin{figure}[h]
  \centering
    \includegraphics[scale=1]{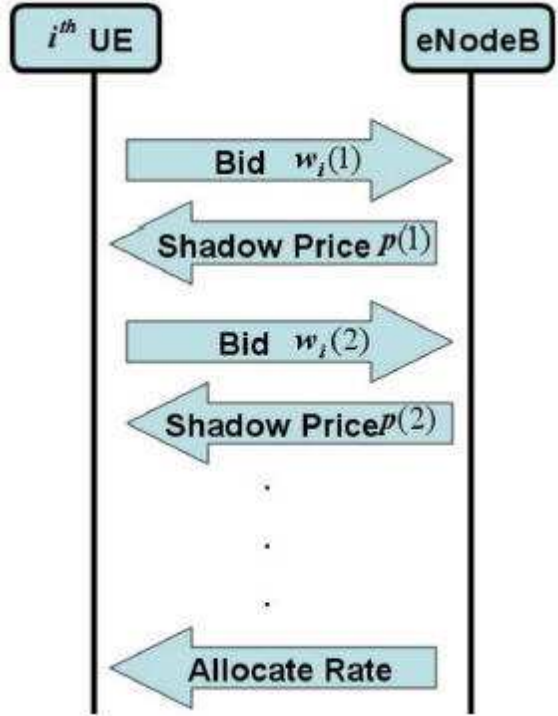}
      \caption{Transmission of Single Carrier with Single Utility per User}
      \label{fig:tx1}
\end{figure}
%%%%%%%%%%%%%%%%%%%%
The implementation of optimization problem using non-linear equation solution:
\begin{itemize}
 \item The solution $r_i$ of the optimization problem $r_{i}(n) = \arg \underset{r_i}\max \Big(\log U_i(r_i) - p(n)r_i\Big)$ in flow chart  in Figure  \ref{fig:UEalgoptimal},  is the value of $r_i$ that solves equation $\frac{\partial \log U_i(r_i)}{\partial r_i} = p(n)$. 
 
 In MATLAB:
  \begin{lstlisting}
dy_sig(i) = a(i).*m(i)./((1+m(i)).*(1-d(i).*(1+m(i))));
dy_log(i) = k(i)./((1+k(i).*x).*log(1+k(i).*x));

\end{lstlisting}
 \item It is the intersection of the horizontal line $y = p(n)$ with the curve $y = \frac{\partial \log U_i(r_i)}{\partial r_i}$ which is calculated in the $i^{th}$ user. 
 
 In MATLAB:
\begin{lstlisting}
soln(i) = fzero(@(x) utility_UE(x,ii,pp),[.001 1000]);

\end{lstlisting}
%\item In C\+\+ implementation, we use bisection method for finding the roots of non-linear equation. This link that has a pseudo-code for it \hyperref{http://en.wikipedia.org/wiki/Bisection_method}
\end{itemize}

%%%%%%%%%%%%%%%%%%%%%%%%%
 \section{Robust Optimal Resource Allocation}\label{sec:robust}
In this section, we present our robust algorithm to ensure the rate allocation algorithms in the flow chart  in Figure \ref{fig:UEalgoptimal} converges for all values of the base station total rate $R$.
%%%%%%%%%%%%%%%%%%%%%%%%%
\subsection{System Model of Robust Resource Allocation}
Similar to Section \ref{subsec:modelsingle}.
%%%%%%%%%%%%%%%%%%%%%%%%
\subsection{Fluctuation Decay Function}
In this section, we show our robust algorithm to ensure the rate allocation algorithms in flow chart Figure \ref{fig:UEalgoptimal} converges for all values of the base station total rate $R$. Our algorithm allocate rates coincide with the Algorithm in flow chart  in Figure \ref{fig:UEalgoptimal} and  in Figure \ref{fig:base stationalgoptimal} for $\sum_{i=1}^{M}b_{i}>R$. For $\sum_{i=1}^{M}b_{i} \ll R$, our algorithm avoids the fluctuation in the non-convergent region discussed in the previous section. This is achieved by: 
\begin{itemize}
 \item adding a convergence measure $\Delta w(n)$ that senses the fluctuation in the bids $w_i$. 
  \item In case of fluctuation, our algorithm decreases the step size between the current and the previous bid $w_i(n) -w_i(n-1)$ for every user $i$ using \textit{fluctuation decay function}. 
\end{itemize}
The fluctuation decay function could be in the following forms:
\begin{itemize}
\item \textit{Exponential function}: It takes the form $\Delta w(n) = l_1 e^{-\frac{n}{l_2}}$.
\item \textit{Rational function}: It takes the form $\Delta w(n) = \frac{l_3}{n}$.
\end{itemize}
where $l_1, l_2, l_3$ can be adjusted to change the rate of decay of the bids $w_i$. 

The new addition in MATLAB with the fluctuation decay function is 
 \begin{lstlisting}
 if abs(w_old(i)-w(i)) > (5.* exp(-0.1*time))%(10 ./ time)
  w(i) = w_old(i) + (5.* exp(-0.1*time)) .* sign(w(i)-w_old(i));
end

\end{lstlisting}
%%%%%%%%%%%%%%%%%%%%%%%%%%%%%%%%%%%
\begin{rem} 
The fluctuation decay function can be included in user Algorithm or base station Algorithm.
\end{rem}
%%%%%%%%%%%%%%%%%%%%%%%%%
\subsection{Algorithm of Robust Optimal Resource Allocation}
  \begin{figure}[h]
  \centering
    \includegraphics[scale=1]{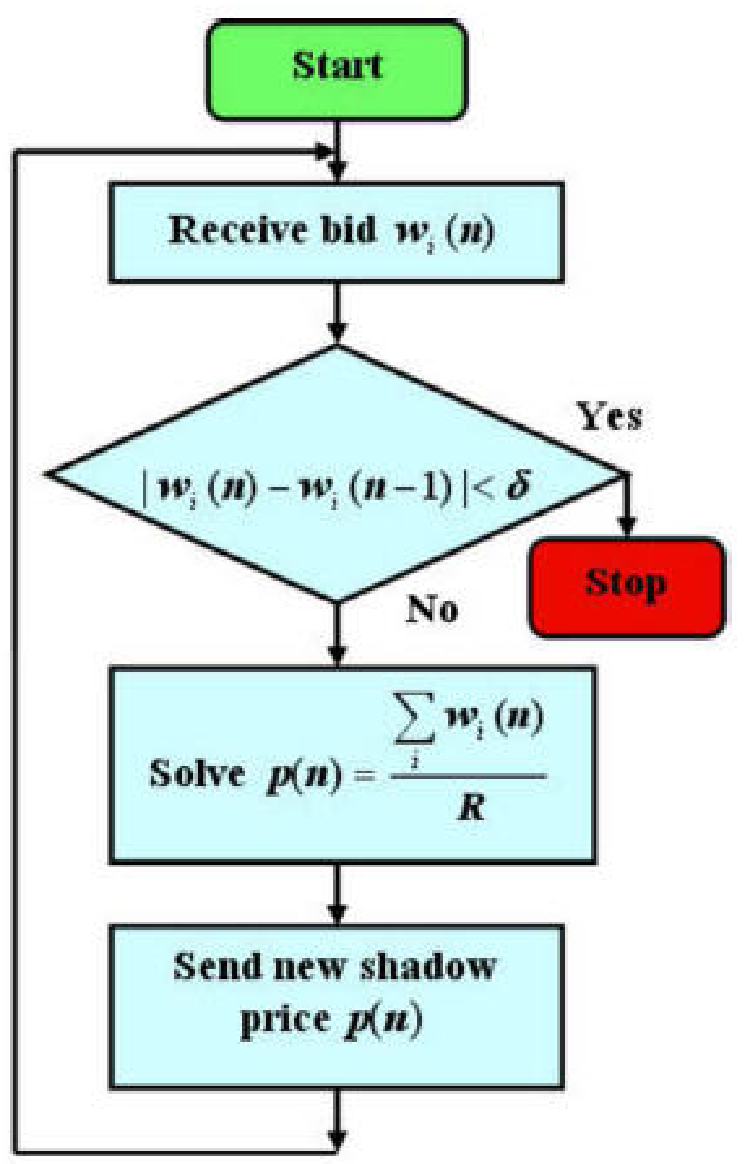}
      \caption{Robust Base Station Algorithm}
      \label{fig:base stationalgrobust}
\end{figure}
%%%%%%%%%%%%%%%%%%%%%
  \begin{figure}[h]
  \centering
    \includegraphics[scale=1.3]{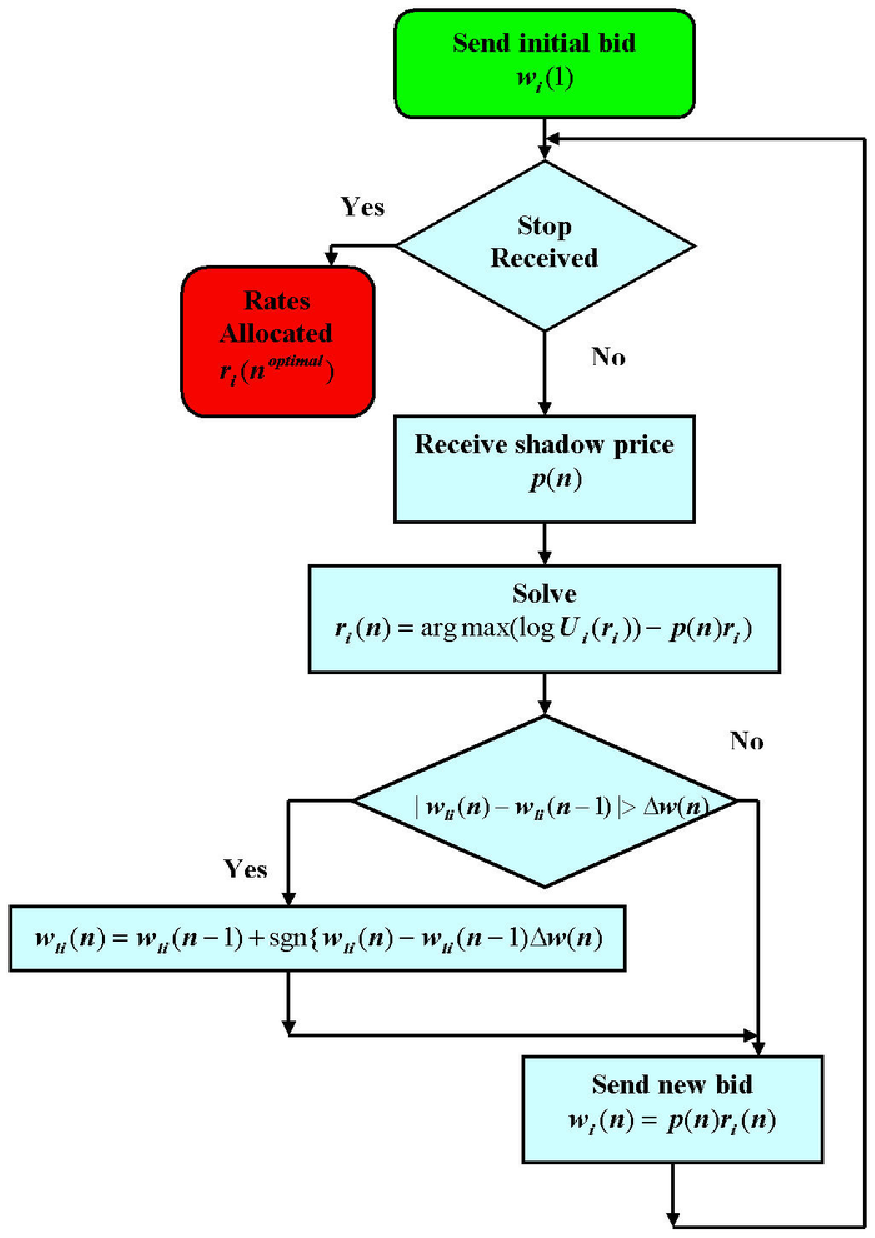}
      \caption{Robust User Algorithm}
      \label{fig:UEalgrobust}
\end{figure}
%%%%%%%%%%%%%%%%%%%%
The algorithm is divided into an user algorithm shown in Figure \ref{fig:UEalgrobust} and an base station algorithm shown in Figure \ref{fig:base stationalgrobust}.  

Flow Chart Description:
\begin{itemize}
 \item Each user starts with an initial bid $w_i(1)$ which is transmitted to the associated base station. 
 
 In MATLAB
 \begin{lstlisting}
 % Initial Bids
w = [10 10 10 10 10 10];

\end{lstlisting}
 \item The base station evaluates the difference between the received bid $w_i(n)$ and the previously received bid $w_i(n-1)$ and exits if it is less than a threshold $\delta$. 
 
 In MATLAB
\begin{lstlisting}
while  (delta > 0.001) %(time<80)%(
    :
    :
    :
    :
    delta = max(abs(w-w_old))
end

\end{lstlisting}
 \item Lets set $w_i(0) = 0$. If the value is greater than the threshold $\delta$, base station calculates the shadow price $p(n) = \frac{\sum_{i=1}^{M}w_i(n)}{R}$ and sends that value to all the users. 
 
 In MATLAB
\begin{lstlisting}
function [p] = base station(w,Rate)
R = Rate;
p = sum(w)/R;

\end{lstlisting}
 \item Each user receives the shadow price to solve for the rate $r_i$ that maximizes $\log U_i(r_i) - p(n)r_i$.  
 
 In MATLAB
 \begin{lstlisting}
 for i = 1: length(a)
    y(i) = log(c(i).*(1./(1+exp(-a(i).*(x-b(i))))-d(i)));
    y(length(a)+i) = log(log(k(i).*x+1)./(1+ log(k(i).*100+1)));
end
for i = 1: 2*length(a)
    dy(i) = diff(y(i),x);
end
:
:
S(i) = dy(i)-p(time);
soln(i,:) = double(solve(S(i)));
:
:

\end{lstlisting}
 \item That rate is used to calculate the new bid $w_i(n)=p(n) r_{i}(n)$. 
 
 In MATLAB
 \begin{lstlisting}
 w(i) = r_opt(i) * p(time);

\end{lstlisting}
\item If the step size between the current and the previous bid $|w_i(n) -w_i(n-1)|$ for every user $i$ is greater than $\Delta w(n)$ then use the \textit{fluctuation decay function}. 

In MATLAB:
\begin{lstlisting}
if abs(w_old(i)-w(i)) > (5.* exp(-0.1*time))%(10 ./ time)
  w(i) = w_old(i) + (5.* exp(-0.1*time)) .* sign(w(i)-w_old(i));
end

\end{lstlisting}
 \item Each user sends its new bid $w_i(n)$ to the base station. This process is repeated until $|w_i(n) -w_i(n-1)|$ is less than the threshold $\delta$. 
 
 In MATLAB:
 \begin{lstlisting}
while  (delta > 0.001) %(time<80)%(
    :
    :
    :
    :
    delta = max(abs(w-w_old))
end

\end{lstlisting}
\end{itemize}
%%%%%%%%%%%%%%%%%%%%

\lstset{ %
language=C++,
basicstyle=\footnotesize,
breaklines=true,
captionpos=b,
}

%\include{chapter-appendix1}

%\include{chapter-appendix2}

%\include{chapter-appendix3}

%%%%%%%%%%%%%%%%%%%%%%%%%%%%%%%%%%%%%%%%%%%%%%%%%%%%%%%%%%%%%%%%%%%%%%
% Generate the bibliography.					     %
%%%%%%%%%%%%%%%%%%%%%%%%%%%%%%%%%%%%%%%%%%%%%%%%%%%%%%%%%%%%%%%%%%%%%%
%								     %
% NOTE: For master's theses and reports, NOTHING is permitted to     %
%	come between the bibliography and the vita. The command      %
%	to generate the index (if used) MUST be moved to before      %
%	this section.						     %
%								     %
%\nocite{*}      % This command causes all items in the 		     %
                % bibliographic database to be added to 	     %
                % the bibliography, even if they are not 	     %
                % explicitly cited in the text. 		     %
		%						     %
%\bibliographystyle{plain}  % Here the bibliography 		     %
%\bibliography{pubs, IEEEabrv,PhD}        % is inserted.			     %
%\index{Bibliography@\emph{Bibliography}}%			     %
%%%%%%%%%%%%%%%%%%%%%%%%%%%%%%%%%%%%%%%%%%%%%%%%%%%%%%%%%%%%%%%%%%%%%%

\include{bibliography}
\markright{References\hfill}
\newpage
\bibliographystyle{ieeetr}
\renewcommand{\bibname}{References}
\bibliography{IEEEabrv,PhD}

\begin{thebibliography}{100}

\bibitem{AbdelhadiICNC2014}
A.~Abdelhadi and C.~Clancy, ``{A Utility Proportional Fairness Approach for
  Resource Allocation in 4G-LTE},'' in {\em IEEE International Conference on
  Computing, Networking, and Communications (ICNC), CNC Workshop}, 2014.

\bibitem{AbdelhadiPIMRC2013}
A.~Abdelhadi and C.~Clancy, ``{A Robust Optimal Rate Allocation Algorithm and
  Pricing Policy for Hybrid Traffic in 4G-LTE},'' in {\em IEEE nternational
  Symposium on Personal, Indoor, and Mobile Radio Communications (PIMRC)},
  2013.

\bibitem{Ghorbonzadeh_book1_chapter3}
M.~Ghorbanzadeh, A.~Abdelhadi, and C.~Clancy, ``Centralized resource
  allocation,'' in {\em Cellular Communications Systems in Congested
  Environments}, pp.~37--60, Springer, 2017.

\bibitem{Ghorbonzadeh_book1_chapter4}
M.~Ghorbanzadeh, A.~Abdelhadi, and C.~Clancy, ``Distributed resource
  allocation,'' in {\em Cellular Communications Systems in Congested
  Environments}, pp.~61--91, Springer, 2017.

\bibitem{Infonetics}
I.~Research, ``{Mobile VoIP subscribers will near 410 million by 2015; VoLTE
  still a long way off},'' 2010.

\bibitem{Nokia2013}
N.~Solutions and Networks, ``{Enhance mobile networks to deliver 1000 times
  more capacity by 2020},'' 2013.

\bibitem{Zokem2011}
G.~Intelligence, ``{Smartphone users spending more 'face time' on apps than
  voice calls or web browsing},'' 2011.

\bibitem{Nokia2011}
N.~S. Networks, ``{Understanding Smartphone Behavior in the Network},'' 2011.

\bibitem{QoE_paper}
S.~Qaiyum, I.~A. Aziz, and J.~B. Jaafar, ``Analysis of big data and
  quality-of-experience in high-density wireless network,'' in {\em 2016 3rd
  International Conference on Computer and Information Sciences (ICCOINS)},
  pp.~287--292, Aug 2016.

\bibitem{QoS_3GPP}
H.~Ekstrom, ``{QoS control in the 3GPP evolved packet system},'' 2009.

\bibitem{Ekstrom2006}
H.~Ekstrom, A.~Furuskar, J.~Karlsson, M.~Meyer, S.~Parkvall, J.~Torsner, and
  M.~Wahlqvist, ``Technical solutions for the {3G} long-term evolution,''
  vol.~44, pp.~38 -- 45, Mar. 2006.

\bibitem{Ghorbonzadeh_book1_chapter1}
M.~Ghorbanzadeh, A.~Abdelhadi, and C.~Clancy, ``Quality of service in
  communication systems,'' in {\em Cellular Communications Systems in Congested
  Environments}, pp.~1--20, Springer, 2017.

\bibitem{Ghosh2011}
A.~Ghosh and R.~Ratasuk, ``{Essentials of LTE and LTE-A},'' 2011.

\bibitem{Piro2010}
G.~Piro, L.~Grieco, G.~Boggia, and P.~Camarda, ``{A two-level scheduling
  algorithm for QoS support in the downlink of LTE cellular networks},'' in
  {\em Wireless Conference (EW)}, 2010.

\bibitem{Monghal2008}
G.~Monghal, K.~Pedersen, I.~Kovacs, and P.~Mogensen, ``{QoS Oriented Time and
  Frequency Domain Packet Schedulers for The UTRAN Long Term Evolution},'' in
  {\em IEEE Vehicular Technology Conference (VTC)}, 2008.

\bibitem{Soldani2011}
D.~Soldani, H.~X. Jun, and B.~Luck, ``{Strategies for Mobile Broadband Growth:
  Traffic Segmentation for Better Customer Experience},'' in {\em IEEE
  Vehicular Technology Conference (VTC)}, 2011.

\bibitem{Hujun2006}
H.~Y. and S.~Alamouti, ``{OFDMA: A Broadband Wireless Access Technology},'' in
  {\em IEEE Sarnoff Symposium}, 2006.

\bibitem{Larmo2009}
A.~Larmo, M.~Lindstrom, M.~Meyer, G.~Pelletier, J.~Torsner, and H.~Wiemann,
  ``{The LTE link-layer design},'' 2009.

\bibitem{ciochina2010}
C.~Ciochina and H.~Sari, ``{A review of OFDMA and single-carrier FDMA},'' in
  {\em Wireless Conference (EW)}, 2010.

\bibitem{Ali2011}
S.~Ali and M.~Zeeshan, ``{A Delay-Scheduler Coupled Game Theoretic Resource
  Allocation Scheme for LTE Networks},'' in {\em Frontiers of Information
  Technology (FIT)}, 2011.

\bibitem{Fudenberg1991}
D.~Fudenberg and J.~Tirole, ``{Nash equilibrium: multiple Nash equilibria,
  focal points, and Pareto optimality},'' in {\em MIT Press}, 1991.

\bibitem{Ranjan2011}
P.~Ranjan, K.~Sokol, and H.~Pan, ``{Settling for Less - a QoS Compromise
  Mechanism for Opportunistic Mobile Networks},'' in {\em SIGMETRICS
  Performance Evaluation}, 2011.

\bibitem{Johari2011}
R.~Johari and J.~Tsitsiklis, ``{Parameterized Supply Function Bidding:
  Equilibrium and Efficiency},'' 2011.

\bibitem{Chung2010}
L.~Chung, ``Energy efficiency of qos routing in multi-hop wireless networks,''
  in {\em IEEE International Conference on Electro/Information Technology
  (EIT)}, 2010.

\bibitem{3GPP}
G.~T.~. V9.0.0, ``Further advancements for e-utra physical layer aspects,''
  2012.

\bibitem{phy_channels_modulation}

\newblock {3GPP Technical Report 36.211, `Physical Channels and Modulation',
  www.3gpp.org}.

\bibitem{phy_procedures}
{3GPP Technical Report 36.213, `Physical Layer Procedures', www.3gpp.org}.

\bibitem{Le2013conf}
L.~B. Le, E.~Hossain, D.~Niyato, and D.~I. Kim, ``Mobility-aware admission
  control with qos guarantees in ofdma femtocell networks,'' in {\em 2013 IEEE
  International Conference on Communications (ICC)}, pp.~2217--2222, June 2013.

\bibitem{Le2013}
L.~B. Le, D.~Niyato, E.~Hossain, D.~I. Kim, and D.~T. Hoang, ``{QoS-Aware and
  Energy-Efficient Resource Management in OFDMA Femtocells},'' {\em IEEE
  Transactions on Wireless Communications}, vol.~12, pp.~180--194, January
  2013.

\bibitem{Andrews2007}
J.~Andrews, A.~Ghosh, and R.~Muhamed, ``Fundamenytals of wimax: Understanding
  broadband wireless netwroking,'' 2007.

\bibitem{Alasti2010}
M.~Alasti, B.~Neekzad, J.~H., and R.~Vannithamby, ``{Quality of service in
  WiMAX and LTE networks [Topics in Wireless Communications]},'' 2010.

\bibitem{Niyato2007}
D.~Niyato and E.~Hossain, ``{WIRELESS BROADBAND ACCESS: WIMAX AND BEYOND -
  Integration of WiMAX and WiFi: Optimal Pricing for Bandwidth Sharing},'' {\em
  IEEE Communications Magazine}, vol.~45, pp.~140--146, May 2007.

\bibitem{IXIACOM2010}
IXIACOM, ``{Quality of Service (QoS) and Policy Management in Mobile Data
  Networks},'' 2010.

\bibitem{FCC2010}
{Federal Communications Commission}, ``{Mobile Broadband: The Benefits of
  Additional Spectrum},'' 2010.

\bibitem{Li2003}
F.~Li, ``{Quality of Service, Traffic Conditioning, and Resource Management in
  Universal Mobile Teleccomunication System (UMTS)},'' 2003.

\bibitem{ETSI2012}
{European Telecommunications Standards Institute}, ``{UMTS; LTE; UTRA;
  E-UTRA;EPC; UE conformance specification for UE positioning; Part 1:
  Conformance test specification},'' 2012.

\bibitem{ETSI2016}
{European Telecommunications Standards Institute}, ``{UMTS; UTRA; General
  description; Stage 2},'' 2016.

\bibitem{Ahmed2004}
N.~Ahmed and H.~Yan, ``{Access control for MPEG video applications using neural
  network and simulated annealing},'' in {\em Mathematical Problems in
  Engineering}, 2004.

\bibitem{Tournier2005}
J.~Tournier, J.~Babau, and V.~Olive, ``{Qinna, a Component-based QoS
  Architecture},'' in {\em Proceedings of the 8th International Conference on
  Component-Based Software Engineering}, 2005.

\bibitem{Jung2011}
I.~Jung, I.~J., Y.~Y., H.~E., and H.~Yeom, ``{Enhancing QoS and Energy
  Efficiency of Realtime Network Application on Smartphone Using Cloud
  Computing},'' in {\em IEEE Asia-Pacific Services Computing Conference
  (APSCC)}, 2011.

\bibitem{Tellabs2012}
Tellabs, ``{Quality of Service in the Wireless Backhaul},'' 2012.

\bibitem{Gorbil2011}
G.~Gorbil and I.~Korpeoglu, ``{Supporting QoS traffic at the network layer in
  multi-hop wireless mobile networks},'' in {\em Wireless Communications and
  Mobile Computing Conference (IWCMC)}, 2011.

\bibitem{Stallings2013}
W.~Stallings, ``Data and computer communications,'' in {\em William Stallings
  Books on Computer and Data Communications}, 2013.

\bibitem{Dovrolis2002}
C.~Dovrolis, D.~Stiliadis, and P.~Ramanathan, ``Proportional differentiated
  services: delay differentiation and packet scheduling,'' 2002.

\bibitem{Sali2005}
A.~Sali, A.~Widiawan, S.~Thilakawardana, R.~Tafazolli, and B.~Evans,
  ``Cross-layer design approach for multicast scheduling over satellite
  networks,'' in {\em Wireless Communication Systems, 2005. 2nd International
  Symposium on}, 2005.

\bibitem{Lutz1991}
E.~Lutz, D.~Cygan, M.~Dippold, F.~Dolainsky, and W.~Papke, ``The land mobile
  satellite communication channel-recording, statistics, and channel model,''
  1991.

\bibitem{Perros1994}
H.~Perros and K.~Elsayed, ``Call admission control schemes: A review,'' 1994.

\bibitem{Kbah_ICNC2016}
Z.~Kbah and A.~Abdelhadi, ``Resource allocation in cellular systems for
  applications with random parameters,'' in {\em 2016 International Conference
  on Computing, Networking and Communications (ICNC)}, pp.~1--5, Feb 2016.

\bibitem{Erpek_SysCon_2016}
T.~Erpek, A.~Abdelhadi, and T.~C. Clancy, ``Application-aware resource block
  and power allocation for lte,'' in {\em 2016 Annual IEEE Systems Conference
  (SysCon)}, pp.~1--5, April 2016.

\bibitem{Braden1994}
R.~Braden, ``{Integrated Services in the Internet Architecture: an Overview},''
  1994.

\bibitem{Braden1997}
R.~Braden, ``{Resource ReSerVation Protocol (RSVP) - Version 1 Functional
  Specification},'' 1997.

\bibitem{Blake1998}
S.~Blake, ``{An Architecture for Differentiated Services},'' 1998.

\bibitem{Nichols1999}
K.~Nichols, ``{A Two-Bit Differentiated Services Architecture for the
  Internet},'' 1999.

\bibitem{Nahrstedt1995}
K.~Nahrstedt, ``{The QoS Broker},'' 1995.

\bibitem{Kushner2004}
H.~Kushner and P.~Whiting, ``Convergence of proportional-fair sharing
  algorithms under general conditions,'' 2004.

\bibitem{Andrews2001}
M.~Andrews, K.~Kumaran, K.~Ramanan, A.~Stolyar, P.~Whiting, and R.~Vijayakumar,
  ``Providing quality of service over a shared wireless link,'' 2001.

\bibitem{UtilityFairness}
G.~Tychogiorgos, A.~Gkelias, and K.~Leung, ``Utility proportional fairness in
  wireless networks,'' IEEE International Symposium on Personal, Indoor, and
  Mobile Radio Communications (PIMRC), 2012.

\bibitem{Li2011}
M.~Li, Z.~Chen, and Y.~Tan, ``A maxmin resource allocation approach for
  scalable video delivery over multiuser mimo-ofdm systems,'' in {\em IEEE
  International Symposium on Circuits and Systems (ISCAS)}, 2011.

\bibitem{Prabhu2010}
R.~Prabhu and B.~Daneshrad, ``An energy-efficient water-filling algorithm for
  ofdm systems,'' in {\em IEEE International Conference on Communications
  (ICC)}, 2010.

\bibitem{DBLP:conf/qosip/Harks05}
T.~Harks, ``Utility proportional fair bandwidth allocation: An optimization
  oriented approach,'' in {\em QoS-IP}, 2005.

\bibitem{Nandagopal2000}
T.~Nandagopal, T.~Kim, X.~Gao, and V.~Bharghavan, ``Achieving mac layer
  fairness in wireless packet networks,'' in {\em Proceedings of the 6th annual
  International Conference on Mobile Computing and Networking (Mobicom)}, 2000.

\bibitem{Ghorbonzadeh_book1_chapter2}
M.~Ghorbanzadeh, A.~Abdelhadi, and C.~Clancy, ``Utility functions and radio
  resource allocation,'' in {\em Cellular Communications Systems in Congested
  Environments}, pp.~21--36, Springer, 2017.

\bibitem{kelly98ratecontrol}
F.~Kelly, A.~Maulloo, and D.~Tan, ``Rate control in communication networks:
  shadow prices, proportional fairness and stability,'' in {\em Journal of the
  Operational Research Society}, 1998.

\bibitem{Low99optimizationflow}
S.~Low and D.~Lapsley, ``Optimization flow control, i: Basic algorithm and
  convergence,'' 1999.

\bibitem{Parekh1993}
A.~Parekh and R.~Gallager, ``A generalized processor sharing approach to flow
  control in integrated services networks: the single-node case,'' 1993.

\bibitem{Demers1989}
A.~Demers, S.~Keshav, and S.~Shenker, ``Analysis and simulation of a fair
  queueing algorithm,'' 1989.

\bibitem{RebeccaThesis}
R.~Kurrle, ``Resource allocation for smart phones in 4g lte advanced carrier
  aggregation,'' {Master Thesis}, {Virginia Tech}, 2012.

\bibitem{abdelhadi2015optimal}
A.~Abdelhadi, A.~Khawar, and T.~C. Clancy, ``Optimal downlink power allocation
  in cellular networks,'' {\em Physical Communication}, vol.~17, pp.~1--14,
  2015.

\bibitem{Abdelhadi_context}
A.~Abdelhadi and T.~C. Clancy, ``Optimal context-aware resource allocation in
  cellular networks,'' in {\em 2016 International Conference on Computing,
  Networking and Communications (ICNC)}, pp.~1--5, Feb 2016.

\bibitem{Boyd}
S.~Boyd and L.~Vandenberghe, {\em Introduction to convex optimization with
  engineering applications}.
\newblock Cambridge University Press, 2004.

\bibitem{Ghorbonzadeh_book1_chapter5}
M.~Ghorbanzadeh, A.~Abdelhadi, and C.~Clancy, ``Resource allocation
  architectures traffic and sensitivity analysis,'' in {\em Cellular
  Communications Systems in Congested Environments}, pp.~93--116, Springer,
  2017.

\bibitem{Lee05non-convexoptimization}
J.~Lee, R.~Mazumdar, and N.~Shroff, ``Non-convex optimization and rate control
  for multi-class services in the internet,'' 2005.

\bibitem{DL_PowerAllocation}
J.~Lee, R.~Mazumdar, and N.~Shroff, ``Downlink power allocation for multi-class
  wireless systems,'' 2005.

\bibitem{Abdelhadi_MobiCom2013}
A.~Abdel-Hadi, C.~Clancy, and J.~Mitola, III, ``A resource allocation algorithm
  for users with multiple applications in 4g-lte,'' in {\em Proceedings of the
  1st ACM Workshop on Cognitive Radio Architectures for Broadband}, CRAB '13,
  (New York, NY, USA), pp.~13--20, ACM, 2013.

\bibitem{GhorbanzadehICNC2015}
M.~Ghorbanzadeh, A.~Abdelhadi, and C.~Clancy, ``{A Utility Proportional
  Fairness Approach for Resource Block Allocation in Cellular Networks},'' in
  {\em IEEE International Conference on Computing, Networking and
  Communications (ICNC)}, 2015.

\bibitem{Erpek2015}
T.~Erpek, A.~Abdelhadi, and C.~Clancy, ``{An Optimal Application-Aware Resource
  Block Scheduling in LTE},'' in {\em IEEE International Conference on
  Computing, Networking and Communications (ICNC) Worshop CCS)}, 2015.

\bibitem{Ghorbonzadeh_book1_chapter6}
M.~Ghorbanzadeh, A.~Abdelhadi, and C.~Clancy, ``Radio resource block
  allocation,'' in {\em Cellular Communications Systems in Congested
  Environments}, pp.~117--146, Springer, 2017.

\bibitem{PCAST2012}
P.~C. o. A. o.~S. Executive Office of~the President and T.~(PCAST), ``Realizing
  the full potential of government-held spectrum to spur economic growth,''
  2012.

\bibitem{Wilson2010}
S.~Wilson and T.~Fischetto, ``Coastline population trends in the united states:
  1960 to 2008,'' in {\em U.S. Dept. of Commerce}, 2010.

\bibitem{Richards2010}
M.~Richards, J.~Scheer, and W.~Holm, ``{Principles of Modern Radar},'' 2010.

\bibitem{FCC_5GHz_Radar06}
{Federal Communications Commission (FCC)}, ``In the matter of revision of parts
  2 and 15 of the commission�s rules to permit unlicensed national
  information infrastructure {(U-NII)} devices in the 5 {GH}z band.'' MO\&O, ET
  Docket No. 03-122, June 2006.

\bibitem{FCC2012}
{Federal Communications Commission}, ``{Proposal to Create a Citizen's
  Broadband Service in the 3550-3650 MHz band},'' 2012.

\bibitem{FCC_NBP10}
{Federal Communications Commission (FCC)}, ``Connecting {A}merica: {T}he
  national broadband plan.'' Online, 2010.

\bibitem{NTIA2010}
NTIA, ``An assessment of the near-term viability of accommodating wireless
  broadband systems in the 1675-1710 mhz, 1755-1780 mhz, 3500-3650 mhz,
  4200-4220 mhz and 4380-4400 mhz bands,'' 2010.

\bibitem{NTIA12}
{National Telecommunications and Information Administration (NTIA)}, ``Analysis
  and resolution of {RF} interference to radars operating in the band 2700-2900
  {MH}z from broadband communication transmitters.'' Online, October 2012.

\bibitem{NTIA2014}
C.~M. and D.~R., ``Spectrum occupancy measurements of the 3550-3650 megahertz
  maritime radar band near san diego, california,'' 2014.

\bibitem{DBLP:conf/globecom/TychogiorgosGL11}
G.~Tychogiorgos, A.~Gkelias, and K.~Leung, ``{A New Distributed Optimization
  Framework for Hybrid Adhoc Networks},'' in {\em GLOBECOM Workshops}, 2011.

\bibitem{DBLP:conf/pimrc/TychogiorgosGL11}
G.~Tychogiorgos, A.~Gkelias, and K.~Leung, ``{Towards a Fair Non-convex
  Resource Allocation in Wireless Networks},'' in {\em IEEE International
  Symposium on Personal, Indoor, and Mobile Radio Communications (PIMRC)},
  2011.

\bibitem{Tao2008}
T.~Jiang, L.~Song, and Y.~Zhang, ``Orthogonal frequency division multiple
  access fundamentals and applications,'' in {\em Auerbach Publications}, 2010.

\bibitem{Yuan10CarrierAggregation}
G.~Yuan, X.~Zhang, W.~Wang, and Y.~Yang, ``{Carrier aggregation for
  LTE-advanced mobile communication systems},'' in {\em Communications
  Magazine, IEEE}, vol.~48, pp.~88--93, 2010.

\bibitem{ResourceAllocationConsideration}
Y.~Wang, K.~Pedersen, P.~Mogensen, and T.~Sorensen, ``Resource allocation
  considerations for multi-carrier lte-advanced systems operating in backward
  compatible mode,'' in {\em Personal, Indoor and Mobile Radio Communications,
  2009 IEEE 20th International Symposium on}, pp.~370--374, 2009.

\bibitem{AbdelhadiICNC2015}
A.~Abdelhadi and C.~Clancy, ``{An optimal resource allocation with joint
  carrier aggregation in 4G-LTE},'' in {\em Computing, Networking and
  Communications (ICNC), 2015 International Conference on}, pp.~138--142, Feb
  2015.

\bibitem{Shajaiah_Springer2015}
H.~Shajaiah, A.~Abdelhadi, and T.~C. Clancy, ``An efficient multi-carrier
  resource allocation with user discrimination framework for 5g wireless
  systems,'' {\em {Springer International Journal of Wireless Information
  Networks}}, vol.~22, no.~4, pp.~345--356, 2015.

\bibitem{GhorbanzadehMILCOM2014}
M.~Ghorbanzadeh, A.~Abdelhadi, and C.~Clancy, ``{A Utility Proportional
  Fairness Bandwidth Allocation in Radar-Coexistent Cellular Networks},'' in
  {\em Military Communications Conference (MILCOM)}, 2014.

\bibitem{Abdelhadi_ICNC_2016_network_MIMO}
A.~Abdelhadi and T.~C. Clancy, ``Network {MIMO} with partial cooperation
  between radar and cellular systems,'' in {\em 2016 International Conference
  on Computing, Networking and Communications (ICNC)}, pp.~1--5, Feb 2016.

\bibitem{Ghorbonzadeh_book1_chapter7}
M.~Ghorbanzadeh, A.~Abdelhadi, and C.~Clancy, ``Spectrum-shared resource
  allocation,'' in {\em Cellular Communications Systems in Congested
  Environments}, pp.~147--178, Springer, 2017.

\bibitem{Lackpour2011}
A.~Lackpour, M.~Luddy, and J.~Winters, ``Overview of interference mitigation
  techniques between wimax networks and ground based radar,'' 2011.

\bibitem{Sanders2013}
F.~Sanders, J.~Carrol, G.~Sanders, and R.~Sole, ``Effects of radar interference
  on lte base station receiver performance,'' 2013.

\bibitem{FHH14}
M.~P. Fitz, T.~R. Halford, I.~Hossain, and S.~W. Enserink, ``{Towards
  Simultaneous Radar and Spectral Sensing},'' in {\em IEEE International
  Symposium on Dynamic Spectrum Access Networks (DYSPAN)}, pp.~15--19, April
  2014.

\bibitem{Hossain_radar1}
Z.~Khan, J.~J. Lehtomaki, R.~Vuohtoniemi, E.~Hossain, and L.~A. Dasilva, ``On
  opportunistic spectrum access in radar bands: Lessons learned from
  measurement of weather radar signals,'' {\em IEEE Wireless Communications},
  vol.~23, pp.~40--48, June 2016.

\bibitem{Abdelhadi_ITW2010}
A.~Abdel-Hadi and S.~Vishwanath, ``On multicast interference alignment in
  multihop systems,'' in {\em IEEE Information Theory Workshop 2010 (ITW
  2010)}, 2010.

\bibitem{Jose_INFOCOM2010}
J.~Jose, A.~Abdel-Hadi, P.~Gupta, and S.~Vishwanath, ``On the impact of
  mobility on multicast capacity of wireless networks,'' in {\em INFOCOM, 2010
  Proceedings IEEE}, pp.~1--5, March 2010.

\bibitem{Abdelhadi_IEEE_SysCon2016}
A.~Abdelhadi, F.~Rechia, A.~Narayanan, T.~Teixeira, R.~Lent, D.~Benhaddou,
  H.~Lee, and T.~C. Clancy, ``Position estimation of robotic mobile nodes in
  wireless testbed using geni,'' {\em CoRR}, vol.~abs/1511.08936, 2015.

\bibitem{Hossain_WiFi1}
S.~Chieochan and E.~Hossain, ``{Wireless Fountain Coding with IEEE 802.11e
  Block ACK for Media Streaming in Wireline-cum-WiFi Networks: A Performance
  Study},'' {\em IEEE Transactions on Mobile Computing}, vol.~10,
  pp.~1416--1433, Oct 2011.

\bibitem{Hossain_WiFi2}
S.~Chieochan and E.~Hossain, ``{Network Coding for Unicast in a WiFi Hotspot:
  Promises, Challenges, and Testbed Implementation},'' {\em Comput. Netw.},
  vol.~56, pp.~2963--2980, Aug. 2012.

\bibitem{Kumar_M2M1}
A.~Kumar, A.~Abdelhadi, and T.~C. Clancy, ``A delay efficient multiclass packet
  scheduler for heterogeneous {M2M} uplink,'' {\em IEEE MILCOM}, 2016.

\bibitem{Kumar_M2M2}
A.~Kumar, A.~Abdelhadi, and T.~C. Clancy, ``An online delay efficient packet
  scheduler for {M2M} traffic in industrial automation,'' {\em IEEE Systems
  Conference}, 2016.

\bibitem{Kumar_M2M3}
A.~Kumar, A.~Abdelhadi, and T.~C. Clancy, ``A delay optimal {MAC} and packet
  scheduler for heterogeneous {M2M} uplink,'' {\em CoRR}, vol.~abs/1606.06692,
  2016.

\bibitem{Kumar_M2M4}
A.~Kumar, A.~Abdelhadi, and T.~C. Clancy, ``A delay-optimal packet scheduler
  for m2m uplink,'' {\em IEEE MILCOM}, 2016.

\bibitem{Kumar_M2M6}
A.~Kumar, A.~Abdelhadi, and T.~C. Clancy, ``A delay efficient mac and packet
  scheduler for heterogeneous m2m uplink,'' {\em IEEE GLOBECOM Workshop on
  Internet of Everything (IoE)}, 2016.

\bibitem{Wang_SysCon2016}
Y.~Wang, A.~Abdelhadi, and T.~C. Clancy, ``Optimal power allocation for lte
  users with different modulations,'' in {\em 2016 Annual IEEE Systems
  Conference (SysCon)}, pp.~1--5, April 2016.

\bibitem{Wilson1993}
F.~Wilson, I.~Wakeman, and W.~Smith, ``{Quality of Service Parameters for
  Commercial Application of Video Telephony},'' 1993.

\bibitem{Shenker95fundamentaldesign}
S.~Shenker, ``Fundamental design issues for the future internet,'' 1995.

\bibitem{Wang_ICNC2016}
Y.~Wang and A.~Abdelhadi, ``A {QoS}-based power allocation for cellular users
  with different modulations,'' in {\em 2016 International Conference on
  Computing, Networking and Communications (ICNC)}, pp.~1--5, Feb 2016.

\bibitem{Ghorbanzadeh_ICNC_2015}
M.~Ghorbanzadeh, A.~Abdelhadi, A.~Amanna, J.~Dwyer, and T.~Clancy,
  ``Implementing an optimal rate allocation tuned to the user quality of
  experience,'' in {\em Computing, Networking and Communications (ICNC), 2015
  International Conference on}, pp.~292--297, Feb 2015.

\end{thebibliography}

\end{document}